\documentclass[aps, prl, reprint, longbibliography, superscriptaddress]{revtex4-2}
\usepackage{graphicx}
\usepackage{xcolor}
\usepackage{hyperref}
\usepackage{orcidlink}

\usepackage[utf8]{inputenc}
\usepackage[T1]{fontenc}
\usepackage{mathptmx}
\usepackage{amsmath}
\usepackage{amssymb}
\usepackage{tabularx}
\usepackage{physics}
\usepackage{braket}
\usepackage{times}

\usepackage{bbding}
\usepackage{tikz}
\usetikzlibrary{decorations.markings, arrows.meta, positioning, fit, quantikz2, shapes.geometric}

\begin{document}

\tikzset{
  operator/.append style={rounded corners},
}

\title{Environment-assisted imaginary time evolution in feedback-based quantum algorithms}

\author{J. C. Bellizotti Souza \orcidlink{0000-0001-8149-5575}}
\email{jcbsouza@dac.unicamp.br}
\affiliation{Instituto de F\'\i sica ``Gleb Wataghin'', Universidade Estadual de Campinas, Campinas, SP, Brazil}

\author{M. C. de Oliveira \orcidlink{0000-0003-2251-2632}}
\email{marcos@ifi.unicamp.br}
\affiliation{Instituto de F\'\i sica ``Gleb Wataghin'', Universidade Estadual de Campinas, Campinas, SP, Brazil}

\date{\today}

\begin{abstract}
  The Feedback-based Algorithm for Quantum Optimization (FALQON) prepares ground states
  without classical parameter optimization, but requires deep circuits and can stagnate
  away from the ground state as a consequence of barren plateaus and symmetry protection.
  In order to avoid stagnation, imaginary time evolution (ITE) layers can be incorporated into FALQON layers,
  enhancing FALQON convergence. However, applying a non-unitary ITE layer in a closed quantum system is not possible,
  and approximations must be made to use ITE on real hardware; using an open quantum system formulation
  to incorporate ITE into FALQON represents a better approach.
  We describe ITE as a quantum operation induced by an environment,
  and combine it with the FALQON feedback law.
  In simulations of a frustrated Ising chain (ANNNI) and of MaxCut,
  the resulting algorithm prepares the ground state starting from any initial state,
  where FALQON can fail, and needs fewer layers on average when the initial state is fixed for the two
  investigated problems. The CITE-FALQON advantage strongly depends on
  the system phase for the ANNNI chain, and on the graph weights for the MaxCut problem.
\end{abstract}

\maketitle

\section{Introduction}

Combinatorial optimization lies at the heart at solving
problems from logistics to finance~\cite{Picariello2025a, Soloviev2025a};
however many are NP-hard~\cite{Karp1972a}, therefore a
technique for solving them are of great interest.
Many of the combinatorial problems can be mapped to an
Ising-like Hamiltonian~\cite{Lucas2013a}, and the solution
to the problem is then encoded in the Hamiltonian ground state,
reducing the task to that of finding the ground state of a
many-body Hamiltonian, which one can attempt to find using quantum computers.
Whether quantum computers can outperform classical computers
remains open~\cite{Abbas2024a}.

On digital circuits (gate-based devices) the near-term approach
is variational~\cite{Preskill2018a, Cerezo2021a}, in which
the circuit is parameterized, with the parameters being classically
optimized to minimize the output state's energy; it is assumed that
one can encode the combinatorial problem as a problem Hamiltonian.
A common algorithm is known as Quantum Approximate Optimization Algorithm (QAOA)~\cite{Farhi2014a,Hadfield2019a,Bartschi2020a,Yoshioka2023a,Blekos2024a,Harrigan2021a}.
Optimizing the parameters is itself NP-hard~\cite{Bittel2021a},
prone to barren plateaus~\cite{McClean2018a, Larocca2022a}, and
sensitive to noise~\cite{Wang2021a}. These factors make classical
optimization expensive and the main algorithms bottleneck.
Besides the classical optimization bottleneck, variational algorithms
such as QAOA are sensitive to the problem Hamiltonian symmetries, limiting
its applicability~\cite{Bravyi2020a}.
Alternative algorithms include decoded quantum interferometry~\cite{Jordan2025a}
and quantum-walk optimizers~\cite{Marsh2020a,Banks2024a,Schulz2024a,Luiz2025a,Nzongani2026a}.

A prominent non-variational approach is based on
Quantum Lyapunov control~\cite{Mirrahimi2005a,Beauchard2007a},
in which feedback laws make a function, such as the energy, decrease monotonically.
Its circuit implementation, the
Feedback-based Algorithm for Quantum Optimization
(FALQON)~\cite{Magann2022a,Magann2022b}, builds the circuit
layer by layer, setting each parameter from measurements on the previous state, without classical optimization.
FALQON was used to find the MaxCut solution on a
superconducting processor~\cite{Magann2022a},
and to find the ground state of the Fermi-Hubbard Hamiltonian~\cite{Larsen2024a}
and the anisotropic next-nearest-neighbor Ising (ANNNI) model~\cite{Pexe2024a}.
Its robustness to control errors~\cite{Legnini2025a}
and a noise-assisted variant~\cite{Swain2025a} were also studied.

The downside of avoiding classical optimization is circuit depth;
FALQON requires deeper circuits~\cite{Magann2022a,Magann2022b,Arai2025a,Rattighieri2025a}
and can stall away from the ground state~\cite{Pexe2026a,Pexe2026b,VanLong2025a}.
Although no classical optimization is present, FALQON is still
susceptible to barren plateaus and symmetry protection.
Attempts to reduce the circuit depth include
second-order feedback~\cite{Arai2025a},
time rescaling~\cite{Rattighieri2025a},
counterdiabatic-inspired fields~\cite{Malla2024a},
perturbative updates of earlier layers~\cite{Brady2025a},
measurement-guided initialization~\cite{Rattighieri2026a},
and imaginary time evolution (ITE)~\cite{VanLong2025a}.
On ITE, the factor $\mathrm{e}^{-H\tau}$,
suppresses excited-states components, with the suppressing
dictated by energy gaps. Appending an ITE layer to each FALQON
step (ITE-FALQON) converges with shallower circuits than FALQON on
Fermi-Hubbard Hamiltonians~\cite{VanLong2025a} and when preparing
thermofield double states~\cite{Pexe2026b}. The exponential
suppression factor is non-unitary, and implementing ITE
on quantum computers requires approximations or overhead,
such as unitaries fitted to measured data~\cite{Motta2020a},
variational states~\cite{McArdle2019a},
probabilistic circuits~\cite{Xie2024a},
ancilla-assisted real-time evolution~\cite{Kosugi2022a},
and multicopy protocols~\cite{Schwartzman2026a}.

Although non-unitary operations are not native
to digital circuits, they arise naturally on open
quantum systems. The unitary evolution of the combined
environment and system of interest can be written in the
operator-sum representation, when the environment is traced
out~\cite{Nielsen2010a}, and by eliminating auxiliary
degrees of freedom one obtains non-unitary operators~\cite{Reiter2012a}.
Engineered dissipation can prepare target states~\cite{Kraus2008a,Diehl2008a,Verstraete2009a},
including ground states without variational optimization~\cite{Cubitt2023a,Watts2026a}.
Moreover, renormalizing the state after a non-unitary
operation makes the evolution nonlinear,
as observed in a dissipative superconducting qubit~\cite{Lee2025a}.

In this work we merge ITE and FALQON as an effective quantum operation
arising from the interaction between the system and an environment,
in which $\mathrm{e}^{-H_p\tau}$ enters as an additional operator
in the operator-sum representation;
we name this approach Channeled Imaginary Time FALQON (CITE-FALQON).
CITE-FALQON is based on the environment-induced non-unitary dynamics
instead of the usual unitary gates, removing the complications
associated with implementing ITE using unitary gates.
We test CITE-FALQON on the ANNNI chain~\cite{Selke1988a,Suzuki2013a}
and on MaxCut.
CITE-FALQON successfully prepares the ground state regardless of the initial state,
whereas FALQON fails from random initial states in the ANNNI chain.
From a fixed initial state, CITE-FALQON requires shallower circuits on average,
however the CITE-FALQON advantage depends strongly on the ANNNI phase, and
in MaxCut on the graph weights.
We make no attempt to engineer the interaction between system and environment,
nor study noise effects. The objective of this work is
to assess whether dissipation-induced non-unitary dynamics can improve feedback-based algorithms.

\section{Methods}

FALQON is a constructive protocol inspired by Quantum
Lyapunov control theory~\cite{Magann2022a}. While variational algorithms,
such as QAOA~\cite{Blekos2024a}, optimize a set of parameters
using classical algorithms, FALQON uses a feedback law, building
the circuit iteratively. The circuit is built iteratively
a FALQON layer is applied, state measurements are made,
an additional circuit layer is added based on the measurement,
and the process repeats until convergence.
The measurement is used to stablish the value of a feedback parameter,
such that the cost function monotonically decreases with the
circuit layer.

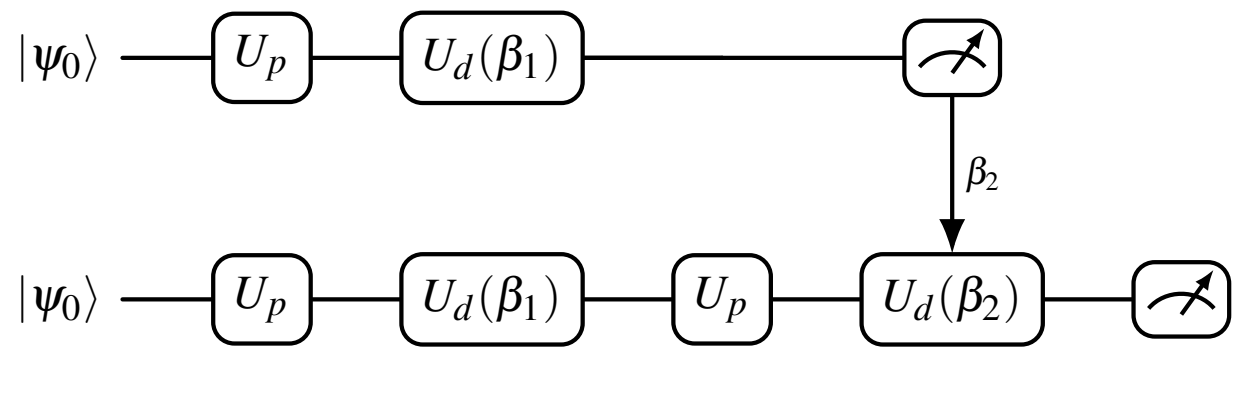
\begin{figure}[!htb]
  \centering
  \tikzset{phase/.append style={scale=2},circlewc/.append style={scale=2}}
  \begin{quantikz}[
      row sep=1cm,
      column sep=0.6cm
    ]
    \lstick{$\ket{\psi_0}$}
    & \gate[1]{U_p}
    & \gate[1]{U_d(\beta_1)}
    & \qw
    & \meter{}
    \arrow[-Latex,d, thick, "\beta_2"{pos=0.5}]
    \\
    \lstick{$\ket{\psi_0}$}
    & \gate[1]{U_p}
    & \gate[1]{U_d(\beta_1)}
    & \gate[1]{U_p}
    & \gate[1]{U_d(\beta_2)}
    & \meter{}
  \end{quantikz}
  \caption{
    Schematic representation of FALQON.
    The measurement after the $k$-th layer determines
    $\beta_{k+1}$ used on the next layer.}
  \label{fig:falqon}
\end{figure}

To prepare the ground state $\ket{\Phi_0}$ of a problem Hamiltonian
$H_p$, one chooses a driver Hamiltonian
$H_d$, such that $[H_p,H_d]\neq0$, typically with an easy-to-prepare ground state;
in this work we fix $H_d=\sum_iX_i$.
The FALQON circuit is then constructed following
\begin{equation}\label{eq:falqon}
  \ket{\psi_k}=\mathrm{e}^{-\mathrm{i}H_d\beta_k\tau}\mathrm{e}^{-\mathrm{i}H_p\tau}\ket{\psi_{k-1}}=U_d(\beta_k)U_p\ket{\psi_{k-1}} \ ,
\end{equation}
where $\tau$ is the fixed evolution time of a FALQON layer, and
$\beta_k$ given by the feedback law
\begin{equation}
  \beta_k=-\mathrm{i}\bra{\psi_{k-1}}[H_d, H_p]\ket{\psi_{k-1}} \ .
\end{equation}
This form of $\beta_k$ ensures that the expectation
value $\bra{\psi_k}H_p\ket{\psi_k}$ decreases monotonically
with $k$ until either (i) the ground state $\ket{\psi_k}=\ket{\Phi_0}$
is prepared (ii) FALQON reaches a barren plateau; the latter case
is undesired and one should be aware of $H_p$ symmetries to
avoid barren plateaus. A schematic representation of FALQON
is shown in Fig.~\ref{fig:falqon}.

In an effort to avoid barren plateaus, \citet{VanLong2025a}
included an ITE step in each FALQON layer,
such that the state at a layer $k$ is given by
\begin{equation}\label{eq:ite-falqon}
  \ket{\psi_k}=\mathcal{N}\mathrm{e}^{-H_p\tau}\mathrm{e}^{-\mathrm{i}H_d\beta_k\tau}\mathrm{e}^{-\mathrm{i}H_p\tau}\ket{\psi_{k-1}} \ ,
\end{equation}
where $\mathcal{N}$ is a normalization constant required
by the non-unitary of the imaginary time evolution.
\citet{VanLong2025a} applied the imaginary time FALQON
to the Fermi-Hubbard Hamiltonian and found that
ITE-FALQON converges at shallower circuits on average than FALQON.
The problem with the approach from Ref.~\cite{VanLong2025a}
is the inclusion of a non-unitary operation in a closed
quantum system. A more appropriate approach is
to consider an open quantum system, where the ITE
is the result of interactions with the environment.

An unitary operation on the combined system,
$\mathcal{H}_s\otimes\mathcal{H}_e$,
becomes non-unitary on $\mathcal{H}_s$ and is described
by the operator-sum representation~\cite{Nielsen2010a},
where a quantum operation $\mathcal{E}(\rho)$ is represented by
\begin{equation}
  \mathcal{E}(\rho)=\sum_kE_k\rho E_k^\dagger \ ,
\end{equation}
and the normalized state after the operation given by
\begin{equation}\label{eq:rho-norm}
  \rho\to \frac{\mathcal{E}(\rho)}{\mathrm{Tr}[\mathcal{E}(\rho)]} \ .
\end{equation}
The normalization requirement introduces non-linearity on quantum channels.

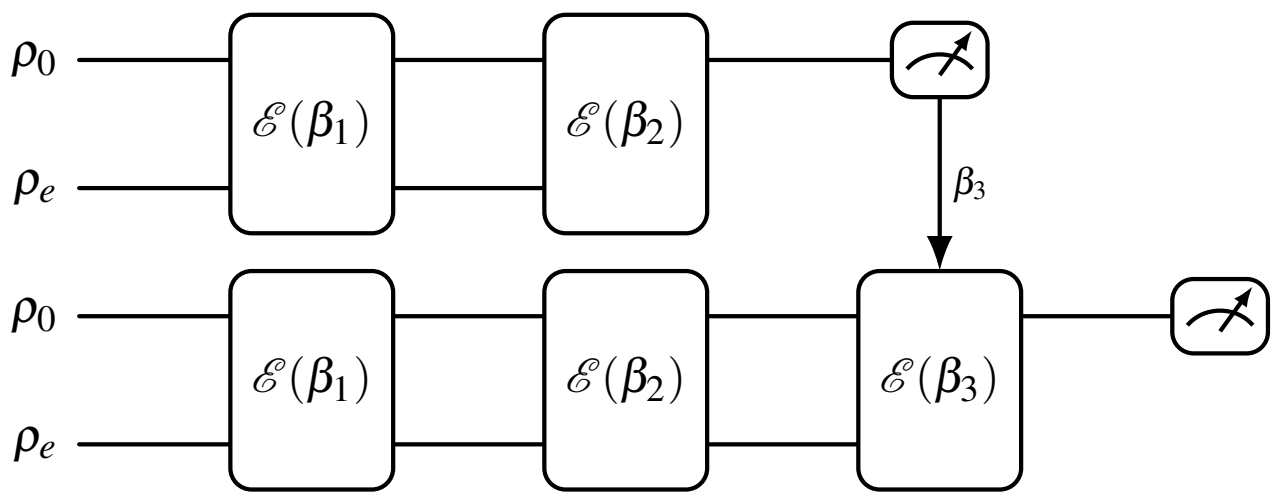
\begin{figure}[!htb]
  \centering
  \begin{quantikz}[
      row sep=0.25cm,
      column sep=1cm
    ]
    \lstick{$\rho_0$}
    & \gate[2]{\mathcal{E}(\beta_1)}
    & \gate[2]{\mathcal{E}(\beta_2)}
    & \meter{}
    \arrow[-Latex, thick, dd, "\beta_3"{pos=0.5}]
    \\
    \lstick{$\rho_e$}
    & &
    \\
    \lstick{$\rho_0$}
    & \gate[2]{\mathcal{E}(\beta_1)}
    & \gate[2]{\mathcal{E}(\beta_2)}
    & \gate[2]{\mathcal{E}(\beta_3)}
    & \meter{}\\
    \lstick{$\rho_e$}
    & & &
  \end{quantikz}
  \caption{
    Schematic representation of CITE-FALQON.
    Each layer is a quantum operation $\mathcal{E}$
    on the system, resulting from interactions with
    the environment. The feedback law is the same
    as in Fig.~\ref{fig:falqon}.
  }
  \label{fig:cite}
\end{figure}

Combining ITE with FALQON using the operator-sum representation, one can write the combined evolution as
\begin{equation}\label{eq:cite-falqon}
  \rho_k=
  \mathrm{e}^{-\mathrm{i}H_d\beta_k\tau}\mathrm{e}^{-\mathrm{i}H_p\tau}\rho_{k-1}
  \mathrm{e}^{\mathrm{i}H_p\tau}\mathrm{e}^{\mathrm{i}H_d\beta_k\tau}
  +\mathrm{e}^{-H_p\tau}\rho_{k-1}\mathrm{e}^{-H_p\tau} \ ,
\end{equation}
followed by the normalization Eq.~\ref{eq:rho-norm}.
ITE is now present as a result of interactions
with environment degrees of freedom, $\mathcal{H}_e$.
The feedback parameter $\beta_k$ is computed in the same manner, $\beta_k=-\mathrm{i}\mathrm{Tr}\left([H_d, H_p]\rho_{k-1}\right)$
\emph{after} $\rho_k$ normalization, Eq.~\ref{eq:rho-norm}. A schematic is shown in Fig.~\ref{fig:cite}.
Interactions with environment
degrees of freedom described by the operator-sum representation can be understood
as a quantum state $\rho$ transmitted through a noisy quantum channel~\cite{Nielsen2010a},
and therefore we name our approach Channeled Imaginary Time FALQON (CITE-FALQON).

We note that, although we refer to our approach as a quantum channel,
quantum channels satisfy $\sum_k E_k^\dagger E_k=1$, which 
is not satisfied here. The naming is based on the use of quantum channels
framework, namely the operator-sum representation.

For all investigated problems we plot the behavior of
$\beta$, approximation ratio $r=\langle H_p\rangle/E_0$
where $E_0$ is the ground state energy,
and success probability $p=\mathrm{Tr}(\rho_k\Pi_{\Phi_0})$,
all as functions of the circuit layer $k$.
We simulate circuits with a maximum depth of $k=1500$,
which we found sufficient to
analyze whether CITE-FALQON achieves convergence faster than FALQON.

\section{ANNNI}

\begin{figure*}[!htb]
  \centering
  \includegraphics[width=\textwidth]{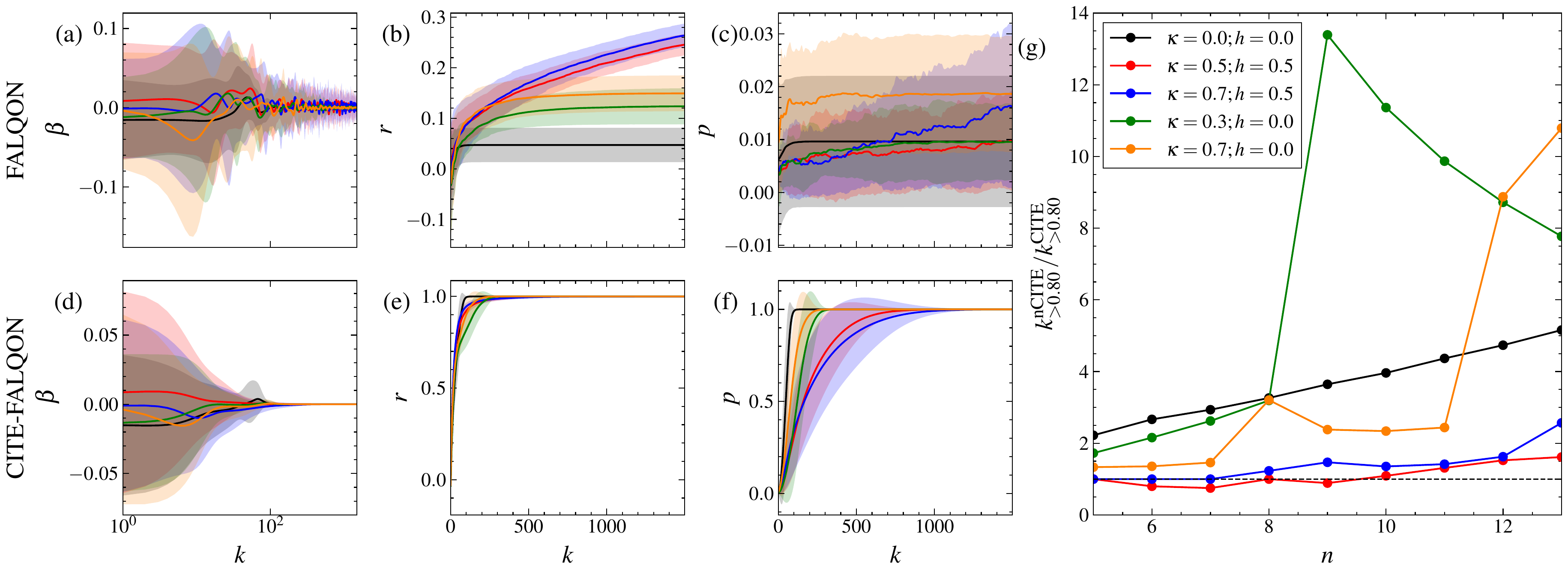}
  \caption{
    Panels (a-c) show FALQON and (d-f) show CITE-FALQON ground state preparation for ANNNI with different
    system parameters $\kappa$ and $h$, with random initial states $\rho_0$.
    (a, d) show $\beta_k$ (b, e) the approximation ratio $r$ and
    (c, f) the success probability $p$, as functions of the layer $k$.
    Panel (g) shows the ratio between the number of required layers
    by FALQON ($k^\mathrm{nCITE}_{>0.80}$) and CITE-FALQON ($k^\mathrm{CITE}_{>0.80}$)
    to reach $p>0.8$ as function of the number of sites $n$,
    with fixed initial state $\rho_0=\ket{+}^{\otimes n}\bra{+}^{\otimes n}$ for
    all pairs $(\kappa, h)$.
  }
  \label{fig:annni-1}
\end{figure*}

As a first application of CITE-FALQON, we prepare the ground state of the ANNNI model.
Anisotropic Next-Nearest Neighbor Ising (ANNNI) consists of a frustrated spin chain
with Hamiltonian
\begin{equation} \label{eq:annni-h}
  H_p=-\sum_{i=1}^{n}\left(Z_iZ_{i+1}-\kappa Z_iZ_{i+2}+hX_i\right) \ ,
\end{equation}
where the first term is the ferromagnetic interaction between
nearest neighbors,
the second term is anti-ferromagnetic
interaction between next-nearest neighbors with strength $\kappa$,
and the third term is the transverse field with strength $h_i$.
The phase diagram as function of $\kappa$ and $h_i=h$ can be found
in Ref.~\cite{Suzuki2013a}.

To study CITE-FALQON convergence on different
phases of the ANNNI model, and how it differs from the convergence
of FALQON, we numerically simulate FALQON and CITE-FALQON circuits
with fixed $\tau=0.2$, and a fixed ANNNI chain size of
$n=8$ spin sites. The set of investigated $\kappa$ and
$h$ is $(\kappa, h)=(0, 0), (0.5, 0.5), (0.7, 0.5), (0.3, 0), (0.7, 0.0)$,
plotted in black, red, blue, green and orange, respectively
on Figs.~\ref{fig:annni-1}(a-f).
We are first interested in how initial
states $\rho_0$ affect convergence;
to address this we investigated
30 random initial states $\rho_0$.

Figure ~\ref{fig:annni-1}(a) shows $\beta$ behavior with
the layer $k$ using FALQON
for the different parameters $\kappa$ and $h$.
In shallower circuits, small $k$,
$\beta$ exhibits large deviations as a consequence of the random initial state,
reflecting the distance \footnote{for any suitable measure of distance}
between the ground state and the initial state, since stronger
effects from the driver Hamiltonian (associated with $|\beta|$) are required to guide
the state towards the ground state.
As $k$ increases, the deviations decrease and
$\beta$ approaches zero for all parameters, with small oscillations
caused by the unitarity of the FALQON evolution. The
diminishing of $\beta$ reflects reduced effects from the driver Hamiltonian.
The behavior is similar across the different investigated parameters.
We note that $\beta\to0$ is \emph{not} sufficient
to guarantee convergence, although it is necessary, as is the case here, where $\beta\approx 0$ while $\rho_k$ differs from the ground state.

Figure~\ref{fig:annni-1}(b) shows FALQON's approximation ratio.
Although $\beta$ is approaching $\beta=0$ at larger $k$,
$\rho_k$ is still distant from the ground state for the investigated
ANNNI parameters.
With $\kappa=h=0$ the approximation ratio does not change,
although $|\beta_k|>0$ indicating that $H_d$
affects $\rho_k$ but it is incapable of bringing it to the ground state.
Systems with $(\kappa, h)=(0.5, 0.5), (0.7, 0.5)$ exhibit
increasing $r$, reaching $r\approx0.24$ with $\kappa=h=0.5$,
and $r\approx0.26$ with $\kappa=0.7, h=0.5$, both with small
deviations at larger $k$. These systems approach the ground state slowly, requiring
$\sim 1500$ layers to improve $r$;
although the approximation ratio exhibited
an increase, the success probability did not, and therefore these systems
were not prepared in the ground state.
Zero field systems, $h=0$, with $\kappa\neq0$, exhibit
an increase in $r$ and then remain constant,
and again FALQON was unable to prepare
the ground state. From the approximation ratio
plot, Fig.~\ref{fig:annni-1}(b), and the strength
of the driver Hamiltonian, $\beta$, Fig.~\ref{fig:annni-1}(a),
we conclude that as the circuit becomes deeper effects from $H_d$
become negligible and FALQON stagnates
without preparing the ground state when \emph{starting
from a random initial state}. Indeed, Fig.~\ref{fig:annni-1}(c)
shows the success probability $p$
as a function of $k$, and for all investigated systems
FALQON was unable to prepare the ground state, with
a success probability of at most $p\approx 2\%$;
we note that the success probability has large relative deviations
with respect to the initial state, revealing FALQON sensitivity
to initial states.
We emphasize that this behavior is specific to random initial
states, and when one initializes the circuit with $\ket{\psi_0}=\ket{+}^{\otimes n}$
FALQON converges to the ANNNI ground state~\cite{Pexe2024a}.

The behavior of CITE-FALQON, Figs.~\ref{fig:annni-1}(d-f),
shares the only similarity that for shallow circuits large
values of $|\beta_k|$ are present, and as the circuit
deepens $\beta_k\to0$. Opposite to FALQON, CITE-FALQON
exhibits $\beta_k\to0$ without oscillations as a consequence
of the non-unitary ITE, an approximation
ratio that rapidly approaches $r=1$, and a success
probability that reached $p=1$ for all investigated systems.

We conclude this section by comparing circuit depth
as a function of the system size $n$ at different ANNNI phases,
by fixing the initial state
$\rho_0=\ket{+}^{\otimes n}\bra{+}^{\otimes n}$.
On Fig.~\ref{fig:annni-1}(g) we plot the ratio between
number of required layers $k$ to achieve $p=80\%$
using FALQON ($k^\mathrm{nCITE}_{>0.8}$) and
CITE-FALQON ($k^\mathrm{CITE}_{>0.8}$). FALQON
prepares the ANNNI ground state with the chosen initial
state~\cite{Pexe2024a}.

For $(\kappa, h)=(0, 0)$, a ferromagnetic phase,
the ratio increases linearly
with the system size $n$, so that FALQON
requires on the order of $0.35n$ times more layers than CITE-FALQON
at larger system sizes.
The systems $(\kappa, h)=(0.5, 0.5)$ and $(0.7, 0.5)$
are paramagnetic, and their ground state is adiabatically connected
to the initial state $\ket{+}^{\otimes n}$. Their ratios are close
to unity for small $n$, but tend to increase with $n$, suggesting
a CITE-FALQON advantage at larger $n$.
The system with $(\kappa, h)=(0.3, 0.0)$,
also a ferromagnetic phase, the ratio increases linearly
for $n\leq 8$, but jumps at $n=9$ to $\sim 14$, and then
decreases staying above unity; we expect the ratio to saturate
at larger $n$.
The last set of parameters, $(\kappa, h)=(0.7, 0)$ is
an antiphase phase, where the ground state is given
by combinations of states of the form $\ket{\uparrow\uparrow\downarrow\downarrow\uparrow\uparrow\cdots}$ with a period of $4$.
A consequence of its periodicity is a different ratio curve
from other phases.
Its ratio exhibits constant behavior for $n<8$,
but at $n=8$ a peak is present, followed by constant
behavior over the interval $8<n<12$, and when $n=12$ is reached
the ratio increases again.
We argue that behavior changes at $n=0\mod 4$ are associated with
matching the antiphase period with the system size, and therefore the ground states
are computational basis states. With $n\neq 0\mod 4$ the ground state cannot be
represented by $\ket{\uparrow\uparrow\downarrow\downarrow\uparrow\uparrow\cdots}$,
instead it is a linear combination of the computational basis states,
and as seen when preparing the paramagnetic phase, states that are combinations
of the computational basis states exhibit a smaller advantage under CITE-FALQON.
We expect this behavior to hold true at larger $n$, outside the investigated
scope.

\section{MaxCut}

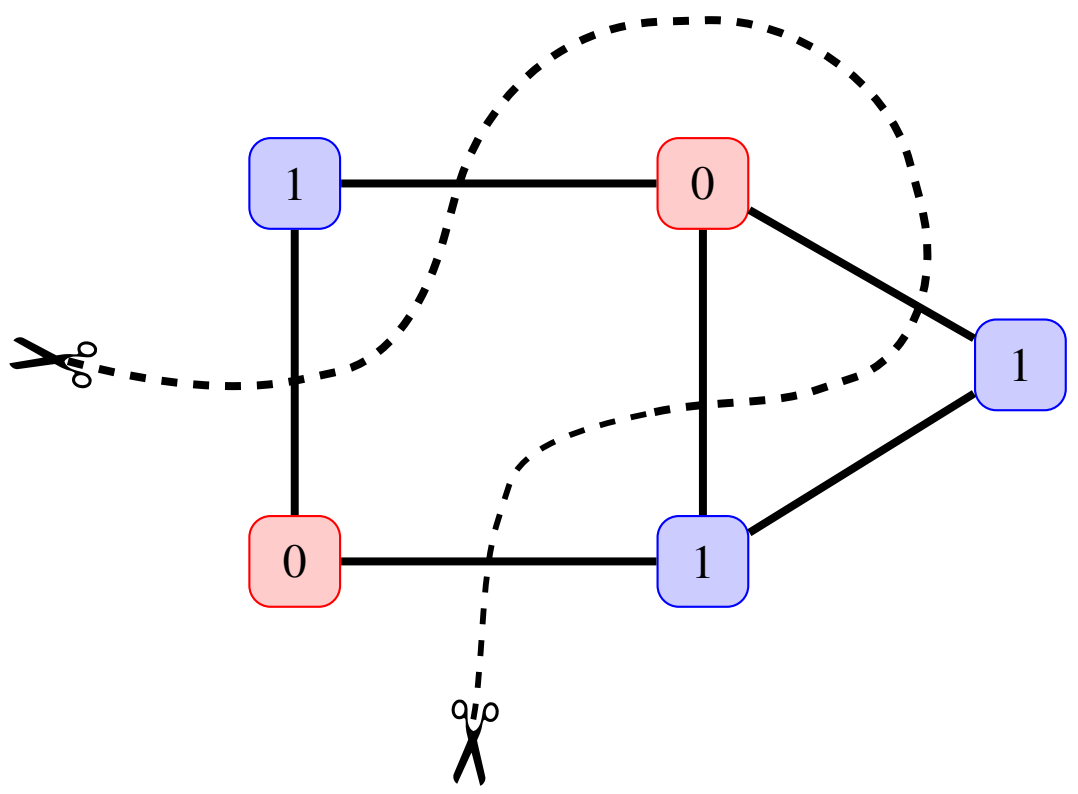
\begin{figure}[!htb]
  \centering
  \begin{tikzpicture}[x=0.5cm,y=0.5cm]
    \node[
      draw,
      shape=rectangle,
      minimum size=0.6cm,
      rounded corners,
      fill = blue!20, draw = blue
    ] (v1) at (3.1,7.7) {1};

    \node[
      draw,
      shape=rectangle,
      minimum size=0.6cm,
      rounded corners,
      fill = red!20, draw = red
    ] (v0a) at (8.5,7.7) {0};

    \node[
      draw,
      shape=rectangle,
      minimum size=0.6cm,
      rounded corners,
      fill = red!20, draw = red
    ] (v0b) at (3.1,2.7) {0};

    \node[
      draw,
      shape=rectangle,
      minimum size=0.6cm,
      rounded corners,
      fill = blue!20, draw = blue
    ] (v1a) at (8.5,2.7) {1};

    \node[
      draw,
      shape=rectangle,
      minimum size=0.6cm,
      rounded corners,
      fill = blue!20, draw = blue
    ] (v1b) at (12.7,5.3) {1};

    \draw[line width=1.5pt]
    (v1) -- (v0a);

    \draw[line width=1.5pt]
    (v1) -- (v0b);

    \draw[line width=1.5pt]
    (v0b) -- (v1a);

    \draw[line width=1.5pt]
    (v0a) -- (v1b);

    \draw[line width=1.5pt]
    (v1a) -- (v1b);

    \draw[line width=1.5pt]
    (v0a) -- (v1a);

    \draw[
      line width=1.5pt,
      dashed,
      rounded corners
    ]
    (0.1,5.35)
    .. controls (1.4,5.0) and (2.4,4.95) .. (3.4,5.15)
    .. controls (4.3,5.35) and (4.8,6.0) .. (5.15,7.25)
    .. controls (5.55,8.75) and (6.55,9.8) .. (8.25,9.85)
    .. controls (9.75,9.9) and (10.95,9.0) .. (11.3,7.75)
    .. controls (11.7,6.25) and (11.15,5.35) .. (10.25,5.05)
    .. controls (9.35,4.75) and (8.55,4.85) .. (7.75,4.65);

    \draw[
      line width=1.1pt,
      dashed,
      rounded corners
    ]
    (7.75,4.65)
    .. controls (6.8,4.45) and (6.1,4.2) .. (5.85,3.45)
    .. controls (5.55,2.45) and (5.6,1.25) .. (5.45,0.55);

    \node[
      rotate=170,
    ] at (-0.1,5.375)
         {{\Large \ScissorRight}};

    \node[
      rotate=265,
    ] at (5.45,0.3)
         {{\Large \ScissorRight}};

  \end{tikzpicture}
  \caption{Example of the MaxCut problem. The sets $S_0$ and $S_1$
    are represented by different colors, and the dashed line
    cuts the edges between them, such that the sum of the weights of the
    cut edges is maximized.}
  \label{fig:maxcut}
\end{figure}

\begin{figure}[!htb]
  \centering
  \includegraphics[width=\columnwidth]{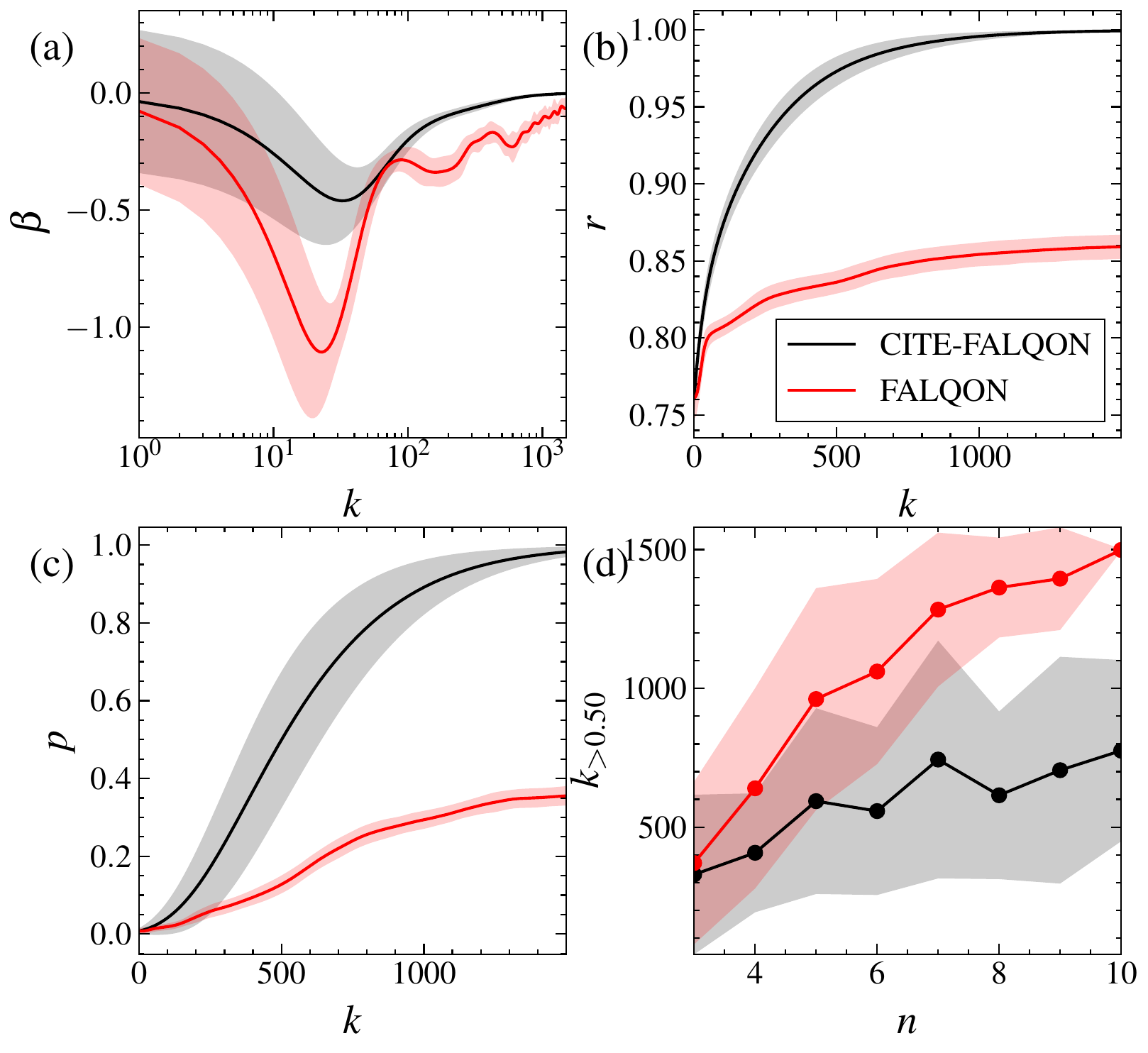}
  \caption{
    CITE-FALQON (black) and FALQON (red) ground state preparation
    for the MaxCut problem.
    (a) $\beta$,
    (b) approximation ratio $r$,
    (c) probability of success $p$
    as functions of the layer $k$
    with random initial states $\rho_0$.
    (d) Number of layers required to
    prepare a state with  $p\geq 0.5$,
    with fixed initial state
    $\rho_0=\ket{+}^{\otimes n}\bra{+}^{\otimes n}$,
    and random graph weights.
  }
  \label{fig:maxcut-plots}
\end{figure}

As a second application, we consider the MaxCut problem.
Given a weighted graph $G=(V, E)$, with weights $w_{ij}$ between
vertices $i$ and $j$ (edge $e_{ij}$), the
MaxCut problem consists of finding two vertex
sets $S_0$ and $S_1$, with $V=S_0\bigcup S_1$, such that
the sum of edge weights between $S_0$ and $S_1$ is maximized,
as illustrated in Fig.~\ref{fig:maxcut}.
The Hamiltonian that encodes the MaxCut solution as its ground
state is given by
\begin{equation} \label{eq:maxcut-h}
  H_p=-\sum w_{ij}\frac{1-Z_iZ_j}{2} \ .
\end{equation}
The basis states $\ket{x_1x_2\cdots x_n}$ in this formulation
are to be interpreted as ``vertex $i$ belongs to partition $S_{x_i}$''.

We start by investigating initial state sensitivity,
the results are present in Fig.~\ref{fig:maxcut-plots}(a-c);
we generated 30 random initial states $\rho_0$ to probe
initial state sensitivity.
The $\beta$ behavior, Fig.~\ref{fig:maxcut-plots}(a), is overall
the same as in Fig.~\ref{fig:annni-1}(a, d). Both FALQON
and CITE-FALQON exhibit larger $|\beta|$ values for shallow circuits,
and approach $\beta=0$ as the circuit deepens.
FALQON exhibits $\beta$ oscillations at deeper circuits, as a consequence
of the unitary evolution, whereas CITE-FALQON exhibits a smooth curve
with $k$ towards $\beta=0$ as the circuit becomes deeper,
a consequence of the non-unitary evolution. Shallow circuits
have $\beta$ values that strongly vary with initial states, but
as the circuit deepens these deviations diminish.
Although one can argue that CITE-FALQON convergence is better
by analyzing $\beta$, this is not sufficient for such claim.

Figure~\ref{fig:maxcut-plots}(b) shows the approximation ratio
for both algorithms. CITE-FALQON approaches $r=1$ faster than FALQON,
reaching $r=1$ at approximately $k=1000$ layers, while FALQON
reaches $r\approx0.86$ at the same circuit depth.
Both algorithms have small deviations across initial states,
indicating that on the MaxCut problem FALQON is robust against
initial states, in contrast to the ANNNI chain.

Lastly we plot the success probability $p$ on
Fig.~\ref{fig:maxcut-plots}(c).
The value of $p$ becomes appreciable only after
$\sim 500$ layers for both algorithms.
For $k>500$
CITE-FALQON increases $p$ with a sigmoid shape until $p=1$,
whereas FALQON increases $p$ more slowly,
at $k=1500$ layers it reaches $p=0.35$ on average.
Both algorithms $p$ values are robust against initial states,
as reflected by the deviations.

We note that on real-world applications
one only has access to $\beta$ and $\langle H_p\rangle$
curves. With CITE-FALQON one could stop increasing
the circuit depth when both $\beta$ and $\langle H_p\rangle$
reach a steady behavior with high confidence of having prepared
the ground state. The same cannot be said about FALQON,
since it is more susceptible to barren plateaus, as in
Fig.~\ref{fig:maxcut-plots}(b, c), where both $r$ and $p$
are close to steady state behavior,
while $\beta$ is close to $\beta=0$ with oscillations,
but $\rho_k$ is not the ground state.

We end this section by investigating convergence as a function
of the graph size $n$, using 30 random graph weights with $n$ vertices,
while the initial state is fixed
$\rho_0=\ket{+}^{\otimes n}\bra{+}^{\otimes n}$, and the
criterion for convergence $p>0.5$ was choses, which
is the usual criterion for optimization tasks, since
the solution state stand out in the distribution.
The results are shown in Fig.~\ref{fig:maxcut-plots}(d).
Since we simulate at most 1500 layers, points with small
deviations close to $k=1500$ are points that failed to converge
to the ground state; this caveat did not apply to analyze
Fig.~\ref{fig:annni-1}(g) because all points converged.
Although both algorithms are robust
against initial states, the same is not true
for robustness against different graphs,
as indicated by large deviations on Fig.~\ref{fig:maxcut-plots}(d).
This is specially true for CITE-FALQON,
where the relative deviation is larger than FALQON's. One
can interpret the large deviations behavior as the ``phase''
of the MaxCut problem, since both algorithms are sensitive to the phase
of Ising-like Hamiltonians,
as seen on the ANNNI system Fig.~\ref{fig:annni-1}(g).
Nevertheless CITE-FALQON requires shallower circuits on average than FALQON,
and converges for all simulated $n$. The same
is not true for FALQON, where signs of nonconvergence can be seen
at larger $n$, where the deviation decreases,
while sensitivity to graph weights increases deviations.
Although CITE-FALQON is strongly affected by graph weights,
it converges with shallower circuits than FALQON on average.

\section{Discussion}

In this work ITE is modeled by interactions
with the environment. This approach is different from other methods
that attempt to implement ITE as unitary evolutions
\cite{Xie2024a, Schwartzman2026a, Kosugi2022a}. Since ITE is
non-unitary these methods can only implement it in approximate form
or probabilistic via measurements,
and often require many ancilla qubits.
Our approach instead takes advantage of the environment to implement ITE
as an effective interaction, described by
the second term in the sum on Eq.~\ref{eq:cite-falqon}.
However this interaction must be carefully engineered;
in this work we did not attempt to engineer such interaction,
since our objective was to assess whether dissipative dynamics could
be used to improve convergence of quantum algorithms.
Engineering this interaction is one of the main challenges
for implementing our approach on real-world hardware.

Environment interaction was modeled as an ITE, however other
effects from the environment, such as noise, may also be present.
We did not investigate noise effects on either
FALQON or CITE-FALQON. Nevertheless we expect noise to
affect both convergence of both algorithms.
As a concrete example, consider the effects of bit-flip noise,
we expect FALQON to exhibit an initial convergence,
but as the circuit becomes deeper, bit-flip effects
accumulate and $r$ changes behavior from strictly increasing to decreasing,
until reaching a steady state, so that FALQON fails to converge.
We expect CITE-FALQON to exhibit a similar initial convergence,
but as the circuit depth increases, bit-flips accumulate and
instead of $r$ changing behavior, as for FALQON,
$r$ saturates at $r<1$. The expected difference in $r$ behavior
is a consequence of the dissipative dynamics of ITE.

Future work should investigate the behavior of CITE-FALQON
in the presence of noise, and explore possible
ways to engineer interactions with the environment so that CITE
can be realized.
Another direction is to investigate other systems, in special
those for which FALQON fails to converge,
such as the Minimum-Spanning-Tree~\cite{Pexe2026a},
and stablish whether CITE-FALQON can prepare ground state,
or if the problem's structure is such that neither FALQON
nor CITE-FALQON is capable of preparing the ground state.

\section{Summary}

In this work we proposed CITE-FALQON, an algorithm for
ground-state preparation based on FALQON, extended with
imaginary time evolution (ITE) using the operator-sum representation.
A method for extending FALQON with ITE was previously described
by \citet{VanLong2025a}, however given the non-unitary nature
of ITE, a better approach is to describe it as an effective
operation $\mathcal{E}$ on a quantum channel, which is the
mechanism underlying CITE-FALQON.
To evaluate whether CITE-FALQON successfully prepares
ground-states, we investigated two systems (i) a condensed
matter application, ANNNI, and (ii) an optimization
application, MaxCut.
We found that CITE-FALQON is robust against different initial
states, preparing the ground-state regardless of the initial
state for both ANNNI and MaxCut.
When investigating the ANNNI system, we found that fixing
the initial state as $\rho_0=\ket{+}^{\otimes n}\bra{+}^{\otimes n}$,
CITE-FALQON requires shallower circuits on average
than FALQON to prepare the ground-states. Furthermore,
the ratio between the number of required layers by the two algorithms
to prepare the ground state strongly depends on the phase of the system.
A similar behavior was observed for MaxCut,
where CITE-FALQON prepares the ground state with shallower circuits on average than
FALQON. Both depend strongly on the graph weights,
as indicated by large deviations across different graph weights;
this dependence can be interpreted as analogous
to the phase dependence observed on ANNNI,
suggesting that different graph weights may
give rise to distinct convergence regimes.
Our approach may be particularly useful for quantum algorithms
operating on open systems, where interactions with the environment
cannot be neglected and may be exploited as part of the computation.

\section*{Data Availability}

All data is available at \url{https://github.com/yuyln/data-cite-falqon}.

\section*{Acknowledgments}

M.C.O. acknowledges partial financial support from the National Institute of Science and
Technology for Applied Quantum Computing through CNPq (Process
No. 408884/2024-0) and from the São Paulo Research Foundation
(FAPESP), through the Center for Research and Innovation on Smart and
Quantum Materials (CRISQuaM, Process No. 2024/00998-6).

J.C.B.S acknowledges financial support from São Paulo Research Foundation
(FAPESP) (Process No. 2025/24799-5) and partial financial support
from the Coordenação de Aperfeiçoamento de Pessoal de Nível Superior - Brasil (CAPES).

\bibliography{bibliography}

\end{document}